\documentclass[twocolumn,prl,floatfix,citeautoscript,nofootinbib,superscriptaddress]{revtex4}
\usepackage{amsbsy}
\usepackage{latexsym,epsfig,graphicx}
\usepackage{dcolumn}
\usepackage{graphicx}
\usepackage{subfigure}
\usepackage{comment}
\usepackage{color}
\usepackage{bm}
\usepackage{mathrsfs}
\usepackage{amsfonts}
\usepackage{amsmath}
\usepackage{color}
\usepackage{amssymb}
\usepackage{xspace}
\usepackage{epstopdf}
\usepackage{tabularx}
\usepackage{longtable}
\usepackage[colorlinks=true, letterpaper=true, pdfstartview=FitV, linkcolor=blue, citecolor=blue, urlcolor=blue]{hyperref}
\usepackage[normalem]{ulem}

\begin{document}

\title{Non-Hermitian Photonics based on Charge-Parity Symmetry}
\author{Junpeng Hou}
\affiliation{Department of Physics, The University of Texas at Dallas, Richardson, Texas
75080-3021, USA}
\author{Zhitong Li}
\affiliation{Department of Electrical and Computer Engineering, The University of Texas
at Dallas, Richardson, Texas 75080-3021, USA}
\author{Qing Gu}
\affiliation{Department of Electrical and Computer Engineering, The University of Texas
at Dallas, Richardson, Texas 75080-3021, USA}
\author{Chuanwei Zhang}
\affiliation{Department of Physics, The University of Texas at Dallas, Richardson, Texas
75080-3021, USA}
\email{chuanwei.zhang@utdallas.edu}

\begin{abstract}
  Parity-time ($\mathcal{PT}$) symmetry, originally conceived for non-Hermitian open quantum systems, has opened an excitingly new avenue for the coherent control of light. By tailoring optical gain and loss in integrated photonic structures, $\mathcal{PT}$ symmetric non-Hermitian photonics has found applications in many fields ranging from single mode lasing to novel topological matters. Here we propose a new paradigm towards non-Hermitian photonics based on the charge-parity ($\mathcal{CP}$) symmetry that has the potential to control the flow of light in an unprecedented way. In particular, we consider continuous dielectric chiral materials, where the charge conjugation and parity symmetries are broken individually, but preserved jointly. Surprisingly, the phase transition between real and imaginary spectra across the exceptional point is accompanied by a dramatic change of the photonic band topology from dielectric to hyperbolic. We showcase broad applications of $\mathcal{CP}$ symmetric photonics such as all-angle polarization-dependent negative refraction materials, enhanced spontaneous emission for laser engineering, and non-Hermitian topological photonics. The $\mathcal{CP}$ symmetry opens an unexplored pathway for studying non-Hermitian photonics without optical gain/loss by connecting two previously distinct material properties: chirality and hyperbolicity, therefore providing a powerful tool for engineering many promising applications in photonics and other related fields.
\end{abstract}

\maketitle

\begin{figure}[t]
  \centering \includegraphics[width=0.48\textwidth]{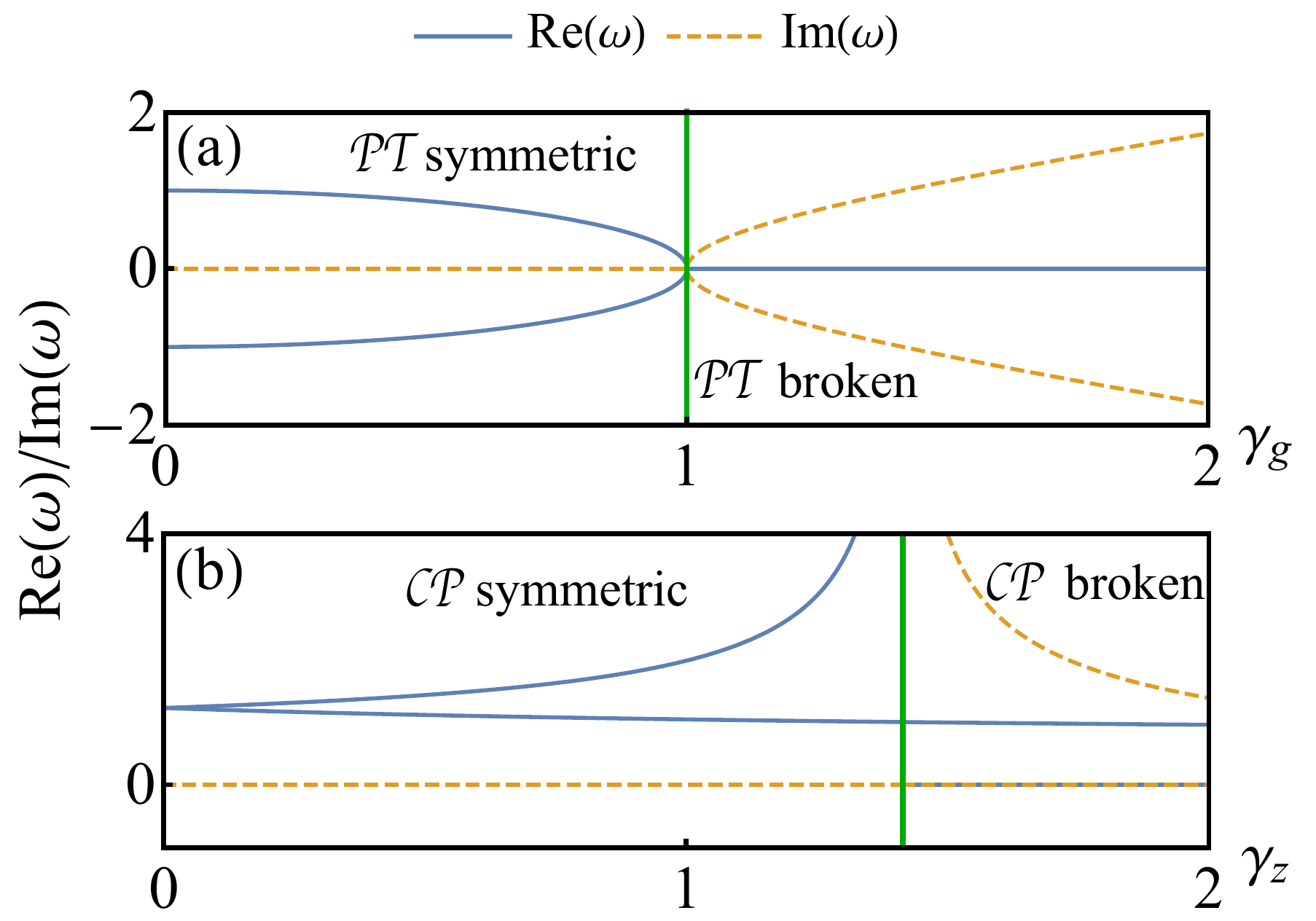}
  \caption{\textbf{Eigenfrequency spectrums for non-Hermitian photonics with (a)} $\mathcal{PT}$ \textbf{and (b)} $\mathcal{CP}$ \textbf{symmetries.} Solid blue and dashed orange curves represent real and imaginary parts of the eigenfrequencies. For the $\mathcal{CP}$ case, each curve is two-fold degenerate with different eigenstates. The green vertical lines indicate the exceptional points, at which the Hamiltonian is defective. The driving terms for the symmetry breaking are gain/loss $\gamma _{g}$ and chiral effect $\gamma _{z}$, respectively. At the exceptional point, a pair of eigenmodes become degenerate at $\omega =0$ for $\mathcal{PT}$ symmetry breaking, while diverge for $\mathcal{CP}$ symmetry breaking. The spectrums for the latter case are computed at $\bm{k}=1$ in a chiral dielectric $\epsilon =2$ and $\mu =1$.}
\label{fig1}
\end{figure}

Originally conceived for open quantum systems \cite{BenderCM1998}, $\mathcal{%
PT}$ symmetry was later introduced to photonics through the analogy between
Schr\"{o}dinger equation and Maxwell equations under paraxial approximation
\cite{FengL2017,LaxM1975}. $\mathcal{PT}$ symmetry allows real
eigenspectrum for a class of non-Hermitian Hamiltonians \cite%
{FengL2017,GanainyR2018}, which support two distinct phases: $\mathcal{PT}$-symmetric (real eigenfrequencies) and $\mathcal{PT}$-broken (both real and complex eigenfrequencies), as illustrated in Fig. \ref{fig1}(a). The phase transition between them is characterized by the exceptional point \cite{HeissWD2004}. In photonics, non-Hermitian Hamiltonian with $\mathcal{PT}$ symmetry can be engineered through tuning optical gain and loss of materials, which provides a powerful tool for shaping the flow of light and yields novel applications in nonlinear optics \cite{WimmerM2015}, lasing\cite{HodaeiH2018,ZhangJ2018}, unidirectional propagation \cite{LinZ2011}, precise sensing \cite{ChenW2018}, topological photonics \cite{WeimannS2017,PanM2018}, \textit{etc}.

The significance of $\mathcal{PT}$ \ symmetric photonics naturally raises
the question whether there is non-Hermitian photonics protected by other
types of symmetries. Note that the $\mathcal{PT}$ \ symmetry for photonics
is based on the paraxial approximation of Maxwell equations, which describe
the multiple-component electromagnetic field and could be non-Hermitian even without optical gain/loss. In this Letter, we propose a new class of
non-Hermitian photonics based on the $\mathcal{CP}$ symmetry of Maxwell
equations, where $\mathcal{C}$ represents charge-conjugation. Similar to $%
\mathcal{PT}$ \ symmetry, there are two distinct phases: $\mathcal{CP}$%
-symmetric with real eigenfrequencies and $\mathcal{CP}$-broken with complex eigenfrequencies. Such $\mathcal{CP}$ symmetric non-Hermitian photonics can
exist, for instance, in a continuous dielectric media with proper chiral
effects. An example is provided in Fig.~\ref{fig1}(b), where the spectrums are two-fold degenerate. The transition between $\mathcal{CP}$-symmetric and -broken phases
can occur at points, lines, or surfaces in either parameter or momentum
space. The transition points are analogous to the exceptional points in $%
\mathcal{PT}$-symmetry in the sense of defective Hamiltonians, therefore we still name them as exceptional points. However, at the exceptional
points, the eigenfrequencies (both real and imaginary parts) diverge (Fig.~\ref{fig1}(b)) for the $\mathcal{CP}$-symmetry, in contrast to the
degeneracy at finite values for the $\mathcal{PT}$-symmetry (Fig. \ref{fig1}%
(a)).

The transition between $\mathcal{CP}$-symmetric and -broken phases is
accompanied by a surprising change of the photonic band topology. In the $\mathcal{CP}$-symmetric phase, the band is still dielectric with elliptical equal frequency surface (EFS), while in the $\mathcal{CP}$-broken phase, the band dispersion becomes hyperbolic with indefinite bands. The hyperbolic
band dispersion is a unique feature of hyperbolic metamaterials (HMMs)
\cite{SmithDR2003,PoddubnyA2013}, which are usually implemented by creating a metal-dielectric composite to achieve metal and dielectric properties in orthogonal directions. HMMs have found great applications in
versatile fields like negative refraction \cite{LiuY2008,ShoaeiM2015},
enhanced spontaneous emission \cite%
{FerrariL2014,GalfskyT2016,LuD2014,FerrariL2017,ChandrasekarR2017},
super-resolution imaging \cite{JacobZ2006,LiuZ2007}, bio-sensing \cite%
{KabashinAV2009}, and topological photonics \cite%
{GaoW2015,ChernRL2017,HouJ2018}. The chirality driven hyperbolic bands
through $\mathcal{CP}$-symmetry breaking provides a new route for realizing
hyperbolic materials in all-dielectric media for the first time, leading to
lossless hyperbolic dielectric materials. The $\mathcal{CP}$-symmetric
physics enables broadly and promisingly technologic applications, and here
we showcase a few examples including all-angle polarization-dependent
negative refraction, enhanced spontaneous emission, and non-Hermitian
topological phases.

\emph{Symmetries of Maxwell equations.} We consider a continuous photonic
medium that can be described by Maxwell equations in the extended
eigenvalue-problem form $H_{P}\psi =\omega H_{M}\psi $ with
\begin{equation}
H_{P}=i\left(
\begin{array}{cc}
0 & \nabla \times  \\
-\nabla \times  & 0%
\end{array}%
\right) ,H_{M}=\left(
\begin{array}{cc}
\epsilon  & i\gamma  \\
-i\gamma  & \mu
\end{array}%
\right) ,\psi =\left(
\begin{array}{c}
\mathbf{E} \\
\mathbf{H}%
\end{array}%
\right) .  \label{Hamiltonian}
\end{equation}%
Here the chiral term $\gamma =\text{Tr}(\gamma )I_{3}/3+N$ \cite%
{ViitanenAJ1994} couples $\bm{E}$ ($\bm{D}$) and $\bm{B}$ ($\bm{H}$%
), $I_{3}$ is the 3$\times $ 3 identity matrix, and $N$ is a symmetric
traceless matrix. For a homogeneous medium, $H(-\bm{k})=-H(\bm{k})$
with $H(\bm{k})=H_{M}^{-1}H_{P}(\bm{k})$ \cite{HouJ2018}, dictating
that eigenmodes $\omega _{\bm{k}}$ and $-\omega _{-\bm{k}}$
represent the same physical state.

The time-reversal symmetry operator is defined as $\mathcal{T}=\sigma
_{z}\otimes I_{3}K$, where the Pauli matrix $\sigma _{i}$ is defined on the $%
\left( \bm{E},\bm{H}\right) $ basis \cite{LuL2014}. Preserving $%
\mathcal{T}$ symmetry requires $\epsilon ^{\ast }=\epsilon $, $\mu ^{\ast
}=\mu $, therefore
gyromagnetic effect or material gain/loss breaks time-reversal symmetry. The
parity symmetry operator $\mathcal{P}=-\sigma _{z}\otimes I_{3}$ satisfies $%
\mathcal{P}H_{P}(\bm{k})\mathcal{P}^{-1}=H_{P}(-\bm{k})$, det$(\mathcal{P})=-1$
\cite{ZeeA2016}, and $\mathcal{P}\left(
\begin{array}{c}
\bm{E} \\
\bm{H}%
\end{array}%
\right) =\left(
\begin{array}{c}
-\bm{E} \\
\bm{H}%
\end{array}%
\right) $ as expected. The Hamiltonian $H(\bm{k})$ has an even parity $%
\mathcal{P}H(-\bm{k})\mathcal{P}^{-1}=H(\bm{k})$ when $\gamma =0$,
and the parity operator changes the sign of the chirality $\gamma $ \cite%
{ViitanenAJ1994}. Because photons are gauge bosons without mass and charge,
the charge-conjugation is defined as $\mathcal{C}=-K$ such that $\mathcal{CPT%
}=1$ \cite{classiccpt}.

In an ideal dielectric (HMM or metal) that only have real permittivity and permeability vectors, these three symmetries are persevered individually. While each of them can be broken explicitly, $\mathcal{CPT}=1$ is always preserved. Note that the chiral term $\gamma =\omega g$ \cite{LeknerJ1996}, thus the signs of eigenfrequency and chiral term are not independent. A chiral inversion operator $\Gamma :\gamma \rightarrow -\gamma $ can be defined, which yields an additional symmetry $(\mathcal{P}\Gamma )H(\bm{k})(\mathcal{P}\Gamma )^{-1}=-H(\bm{k})$, dictating that there are only
two independent non-zero modes. Hereafter we choose both modes with $\Re
(\omega )>0$ $(<0)$ for a given chiral term $\gamma $ ($-\gamma $) \cite%
{HouJ2018}. The above symmetries of Maxwell equations and their consequences
are summarized in Tab.~\ref{tab1}.

\begin{table}[t]
  \begin{tabular}{@{}lll}
  \toprule Symmetry & Eigenmodes & Breaking mechanism(s) \\
  $\mathcal{T}$ & $\omega_{\bm{k}}\to\omega_{-\bm{k}}^*$ & Gain/loss, gyromagnetic \\
  $\mathcal{C}$ & $\omega_{\bm{k}}\to\omega_{\bm{k}}^*$ & Gain/loss, chiral, gyromagnetic \\
  $\mathcal{C}\Gamma$ & $\omega_{\bm{k}}\to\omega_{\bm{k}}^*$ & Gain/loss, gyromagnetic \\
  $\mathcal{P}$ & $\omega_{\bm{k}}\to\omega_{-\bm{k}}$ & Chiral \\
  $\mathcal{P}\Gamma$ & $\omega_{\bm{k}}\to\omega_{-\bm{k}}$ & Cannot be broken
  \end{tabular}
  \caption{\textbf{Symmetries of Maxwell equations.} Some important symmetries, together with their actions on eigenmodes and breaking methods, are listed for $H(\bm{k})$. Besides $\mathcal{CPT}$, two symmetries $H(-\bm{k})=-H(\bm{k})$ and $\mathcal{P}\Gamma $ are always preserved.}
  \label{tab1}
\end{table}

Interesting physics arises when a combination of two symmetries, such as $%
\mathcal{PT}$ or $\mathcal{CP}$, is preserved while each is individually
broken. Here we consider non-Hermitian photonics based on $\mathcal{CP}$
symmetry. Consider the two eigenmodes $\psi _{j,\bm{k}}$ with
eigenfrequencies $\omega _{j,\bm{k}},j=0,1$ of the Maxwell equations.
Under $\mathcal{CP}$ symmetry, $(\mathcal{C}\mathcal{P})H(\bm{k})(%
\mathcal{CP})^{-1}=H(-\bm{k})$, meaning $\omega _{j,\mathbf{-k}}^{\ast }$ is also an eigenfrequency. Notice that the constantly preserved $(\mathcal{P}\Gamma )H(\bm{k})(\mathcal{P}\Gamma )^{-1}=H(-\bm{k})$ symmetry, yields that $\omega _{j,\bm{k}}=\omega _{j,-\bm{k}}$. Note that, such a constraint always applies and it gives rise to the two-fold degeneracy in Fig.~\ref{fig1}(b).

Therefore, in the $\mathcal{CP}$ symmetric regime, the combination of the above two conditions gives $\omega _{j,\bm{k}}=\omega _{j,-\bm{k}}^*=\omega _{j,\bm{k}}^*$. This imposes the reality of eigenfrequency spectrum. In this symmetric regime, the eigenstates also obey the transformation relation given by the $\mathcal{CP}$ symmetry, i.e., $(\mathcal{CP})\psi _{j,\bm{k}}=e^{i\theta_{j}}\psi _{j,-\bm{k}}$. There also exists the $\mathcal{CP}$-broken regimes, where $\mathcal{CP}$ symmetry relates the two states for a given $j$ at the same, instead of opposite, momenta. This leads to the constraints on eigenstates $(\mathcal{CP})\psi _{j,\pm\bm{k}}=e^{i\theta_{j,\pm\bm{k}}}\psi _{j,\pm\bm{k}}$, meaning the eigenfrequencies must be purely imaginary since $\omega _{j,\pm\bm{k}}=-\omega _{j,\pm\bm{k}}^*$.

\begin{figure*}[t]
  \centering \includegraphics[width=0.91\textwidth]{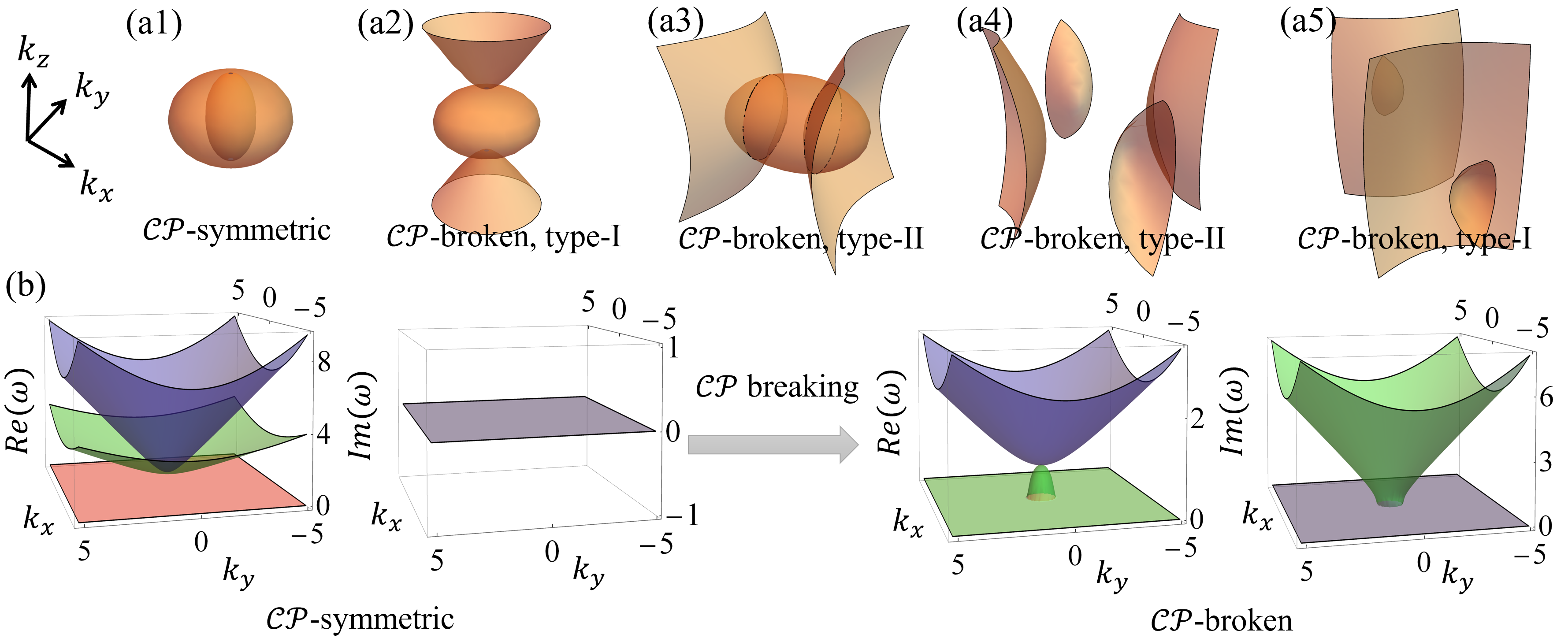}
  \caption{$\mathcal{CP}$ \textbf{breaking and hyperbolicity.} (a) Some examples of EFSs at $\protect\omega =1$ in $\mathcal{CP}$-symmetric (-broken) regions. We define the type-I and type-II HMMs according to det$(H_{M})<0$ and det$(H_{M})>0$ respectively. From left to right, we choose $\protect\gamma =\text{diag}(0,0,1)$, $\protect\gamma =\text{diag}(0,0,2)$, $\protect\gamma =\text{diag}(0,2,2)$, $(\protect\gamma _{x},\protect\gamma _{y},\protect\gamma _{xy},\protect\gamma _{yx})=(1,1,3,3)$ and $(\protect\gamma _{x},\protect\gamma _{y},\protect\gamma _{z},\protect\gamma _{xy},\protect\gamma _{yx})=(1,1,2,3,3)$. (b) Dielectric real bands (a1) change to type-I hyperbolic complex bands (a2) across $\mathcal{CP}$ symmetry breaking in momentum space $k_{z}=1$. For all panels, $\protect\epsilon _{D}=2$ and $\protect\mu _{D}=1$.}
  \label{fig2}
\end{figure*}

$\mathcal{CP}$ \emph{symmetric photonics in a chiral medium}. The chiral
term $\gamma $ breaks $\mathcal{C}$ and $\mathcal{P}$ symmetries
individually, but preserves the combined $\mathcal{CP}$ symmetry, therefore
chiral media provide an excellent platform for exploring non-Hermitian
photonics based on $\mathcal{CP}${\emph{\ }}symmetry. Chirality is
ubiquitous in many different materials, and in photonics chirality gives
rise to unique wave propagations \cite{LeknerJ1996}. However, chiral
effects are usually weak in natural materials. Recently, rapid development
of chiral metamaterials \cite{OhSS2015,WangZ2016,MaX2017} and chiral
plasmonics \cite{HentschelM2017} has opened the door for realizing strong
chiral media in a wide range of frequencies including microwave, terahertz,
infrared and visible frequencies. Besides enhanced circular dichroism and optical activity \cite{OhSS2015,WangZ2016,MaX2017,HentschelM2017}, strong chiral media also exhibits negative refraction for certain incident angles \cite{PendryJB2004,MonzonC2005,ChernR2013,ZhangS2009}.

For better illustration of $\mathcal{CP}$ symmetry effects, we consider an
isotropic dielectric $\epsilon =\epsilon _{D}I_{3},\epsilon _{D}\geq 0$, $%
\mu =\mu _{D}I_{3},\mu _{D}>0$ with only real diagonal chiral term $\gamma =%
\text{diag}(\gamma _{x},\gamma _{y},\gamma _{z})$. Such chiral term breaks $%
\mathcal{C}$, $\mathcal{P}$ respectively, but preserves their combination $%
\mathcal{CP}$ (see Tab.~\ref{tab1}). A simple but instructive case is $%
\gamma =$diag$(0,0,\gamma _{z}>0)$. The EFS can be determined by
\begin{equation}
\epsilon_D\mu_D(k_t^2+k_z^2-\epsilon_D\mu_D\omega^2)^2=
\gamma_z^2(k_z^2-\epsilon_D\mu_D\omega^2)^2,  \label{EFS}
\end{equation}
which has four solutions in general
\begin{equation}
k_{z}=\pm\sqrt{f_{\pm }(\gamma _{z}) k_{t}^{2}+\omega ^{2}\epsilon _{D}\mu
_{D}},
\end{equation}
where $f_{\pm }(\gamma _{z}) =\frac{\sqrt{\epsilon_D\mu_D}}{\pm\gamma_z-%
\sqrt{\epsilon_D\mu_D}}$ and $k_{t}^{2}=k_{x}^{2}+k_{y}^{2}$.

For a small $\gamma _{z}<$ $\sqrt{\epsilon _{D}\mu _{D}}$, $f_{\pm }\left(
\gamma _{z}\right) $ are both negative, therefore the EFS is bounded and
elliptical, which is essentially the same as a dielectric (see Fig.~\ref%
{fig2}(a1)), except that the degeneracy between different polarizations is
lifted. All eigenfrequencies are real and the same at $\pm \bm{k}$. The
eigenmodes with nonzero eigenfrequencies satisfy $(\bm{E}_{\bm{k}},%
\bm{H}_{\bm{k}})_{j}=e^{i\theta _{j}}(-\bm{E}_{-\bm{k}%
}^{\ast },\bm{H}_{-\bm{k}}^{\ast })_{j}$, demonstrating the $%
\mathcal{CP}$-symmetric phase.

At the exceptional point $\gamma _{z}^{c}=\sqrt{\epsilon _{D}\mu _{D}}$, $%
f_{+}\left( \gamma _{z}\right) $ diverges and the Hamiltonian $H(\bm{k})$
is ill-defined because det$(H_{M})=0$. There are only two solutions for $%
f_{-}\left( \gamma _{z}\right) $ with $k_{z}=\pm \sqrt{-k_{t}^{2}/2+%
\omega^{2}\epsilon _{D}\mu _{D}}$. The non-Hermitian Hamiltonian $H(\mathbf{k%
})$ is defective at the exceptional point in the sense that the number of
linearly independent eigenmodes is less than the dimension of the
Hamiltonian, which is different from the defectiveness of $\mathcal{PT}$%
-symmetric Hamiltonians at the exceptional point that have coalesced
eigenstates (i.e., linearly dependent).

Beyond the exceptional point $\gamma _{z}>\gamma _{z}^{c}$, two purely
imaginary eigenmodes appear,
which denotes the $\mathcal{CP}$-broken regime. Meanwhile, $f_{+}\left( \gamma_{z}\right) $ becomes positive, leading to the indefinite (hyperbolic)
bands, as shown in Fig.~\ref{fig2}(a2), which are similar to type-I HMMs. In Fig.~\ref{fig2}(b), we plot the complex band structures at a finite $k_{z}$
across the $\mathcal{CP}$ breaking transition. Before the transition $%
\gamma_{z}<\gamma _{z}^{c}$, the degenerate bands (blue and green) at\ $%
\gamma_{z}=0$ split but remain real in the entire momentum space. The red
plane represents static solutions of Maxwell equations, which are zero for
any chiral term. With increasing $\gamma _{z}$, the lower band is gradually
flattened. Across the exceptional
point $\gamma _{z}^{c}$, the real part of the lower band manifests as a cone
located at the origin with a quadratic band touching with the upper band,
while the rest parts become purely imaginary. Such band dispersion exhibits
an exceptional ring on the $k_{x}$-$k_{y}$ plane (an exceptional cone in 3D
momentum space), where the eigenmodes coalesce to zero eigenfrequency
and null eigenvector.

\begin{figure*}[t]
\centering \includegraphics[width=1\textwidth]{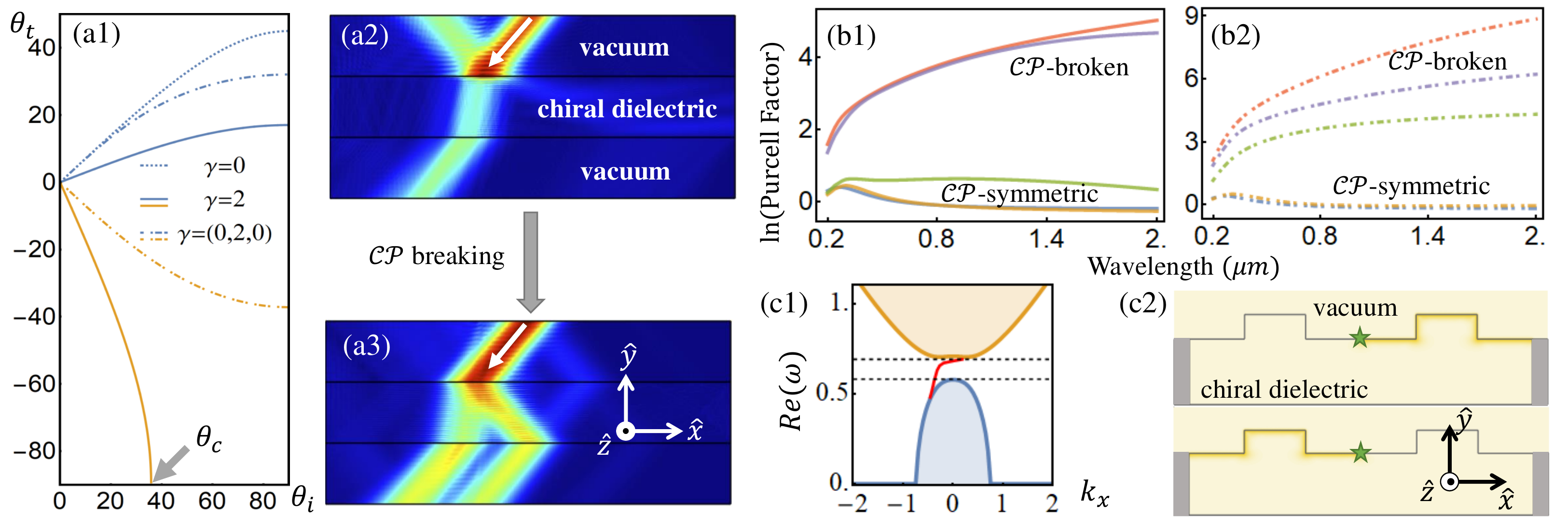}
\caption{\textbf{Applications of} $\mathcal{CP}$\textbf{-symmetric photonics}%
. (a) All-angle and polarization-dependent negative refraction. (a1)
Analytic calculations of a linearly polarized plane wave transmitting on the
boundary between a chiral dielectric and vacuum. The plot shows transmitted
angle $\protect\theta _{t}$ with respect to different incident angle $%
\protect\theta _{i}$. The blue and khaki curves represent right-handed and
left-handed polarizations. (a2-a3) COMSOL simulations for (a1). Incident
angle is fixed at $\protect\theta _{i}=40^{\circ }$ while $\protect\gamma %
=2I_{3}$ and $\protect\gamma =\text{diag}(0,2,0)$ for the upper and lower
panel. The white arrows denote the incident directions. (b) Isotropic
Purcell factor for $\mathcal{CP}$ symmetric dielectric with (b1) type-I and
(b2) type-II hyperbolic dispersions after symmetry breaking. A leap of
Purcell factor is observed across the exceptional point. (c) Topological
physics in the $\mathcal{CP}$-broken phase. (c1) 2D band structure at $%
k_{z}=1$ with open boundary condition along $y$. Two dashed lines give the
band gap 0.58 to 0.69 and the red curve represents chiral surface wave. (c2)
COMSOL simulations for $k_{z}=1$ (upper) and $k_{z}=-1$ (lower). The green
star denotes the location of a line source and the corresponding input
energy is $\protect\omega =0.59$. The grey areas represent absorption
materials. For all panels, $\protect\epsilon _{D}=2$ and $\protect\mu _{D}=1$%
.}
\label{fig3}
\end{figure*}

The existence of multiple chiral terms along different directions provide a
tunable tool for driving $\mathcal{CP}$ breaking transition and engineering
different hyperbolic band dispersions. In the case $\gamma =\text{diag}%
(\gamma _{x},\gamma _{y},\gamma _{z})$, there exists three individual
exceptional points at $\gamma _{l}^{c}=\sqrt{\epsilon _{D}\mu _{D}},l=x,y,z$
along each spatial direction. When all $\gamma _{l}<\sqrt{\epsilon _{D}\mu
_{D}}$, the system remains $\mathcal{CP}$ symmetric and have ellipsoid EFSs.
When one chiral term such as $\gamma _{z}$ exceeds $\gamma _{z}^{c}$, the
system enters the $\mathcal{CP}$-broken phase and exhibits hyperbolic
dispersion, as shown in Figs.~\ref{fig2}(a2) and (b). When $\gamma _{y}$
also exceeds the exceptional point, the hyperbolic dispersion changes from
type-I to type-II (see Fig.~\ref{fig2}(a3)), and the cone-like EFSs lie
along the $k_{x}$ direction, which is perpendicular to the $k_{z}$-$k_{y}$
plane. Interestingly, when all three chiral components exceeds $\gamma _{l}^{c}$%
, the system transitions back to the $\mathcal{CP}$-symmetric phase and the
hyperbolic dispersions disappear. This feature is unique to $\mathcal{CP}
$ symmetric systems, because in $\mathcal{PT}$ symmetric systems, one always ends up in the $\mathcal{PT}$-broken
regime with increasing material gain/loss strength. More exotic hyperbolic band dispersions, which cannot be
realized in metal-dielectric patterned HMMs, can be engineered through $%
\mathcal{CP}$-breaking when non-diagonal chiral terms are involved, as shown
in two examples illustrated in Figs.~\ref{fig2}(a4) and (a5). More details
on the $\mathcal{CP}$-breaking-induced hyperbolicity are presented in
Supplemental Note \cite{Supp}. These results clearly showcase that strong
chiral materials may provide a tunable platform for realizing lossless
hyperbolic materials and significantly broadening the applications of HMMs.

\emph{Applications of} $\mathcal{CP}$ \emph{breaking.} Due to the rise of
hyperbolicity, the non-Hermitian photonics based upon $\mathcal{CP}$
symmetry may have potential applications in a plethora of fields. In the
following, we briefly illustrate three applications covering classical
optics, laser engineering, and topological photonics, and leave the
technical details in Supplemental Notes \cite{Supp}.

\textbf{All-angle polarization-dependent beam splitter}: Birefringence and
negative refraction have long been studied in chiral media and it is known
that there is a critical angle $\theta _{c}$, beyond which one polarization
is totally reflected. The critical angle vanishes in $\mathcal{CP}$-broken
regimes because the negative refraction has been promoted to an all-angle
effect thanks to indefinite bands, as demonstrated by the analytic results
(Fig.~\ref{fig3}(a1)), together with the COMSOL simulation (Figs.~\ref{fig3}%
(a2,a3)). Because of the polarization dependence of the chiral media shown
in Fig.~\ref{fig3}(a), the $\mathcal{CP}$-symmetric photonics
promises an all-angle polarization-dependent beam splitter, a device that is
hard to engineer in either isotropic chiral materials or HMM.

\textbf{Enhanced spontaneous emissions for laser engineering}: Spontaneous
emissions play a crucial role for laser engineering, and have been widely
studied in chiral media, but mainly in the $\mathcal{CP}$-symmetric regime.
The hyperbolic bands in the $\mathcal{CP}$-broken regime can significantly
enhance the spontaneous emissions of a dielectric continuum with both
broader bandwidth and stronger Purcell effect. For simplicity, we assume a
frequency-independent permeability constant and compute the isotropic
Purcell factor in Fig.~\ref{fig3}(b). Upon crossing the exceptional point, a large leap of the Purcell factor is observed.

\textbf{Topological photonics}: Interestingly, the $\mathcal{CP}$ breaking
transition could also be a topological one, leading to a non-trivial
topological phase in the $\mathcal{CP}$-broken phase. The topological
properties of a $\mathcal{CP}$-broken chiral dielectric is similar to that
of a Weyl semimetal but the Weyl point is replaced by a charge-2
triply-degenerate point \cite{HouJ2018,HuH2018} at $\bm{k}=0$ (see
Supplemental Note \cite{Supp} for more details). At a finite $k_{z}$, the
quadratic band touching in Fig.~\ref{fig2}(b) may be lifted by anisotropy of
$\epsilon $ and the projected 2D band structure with chiral surface wave
is plotted in Fig.~\ref{fig3}(c1). Such chiral edge states are confirmed in
COMSOL simulation (Fig.~\ref{fig3}(c2)), where we see the chirality of edge
states is associated with the sign of $k_{z}$ due to the breaking of $%
\mathcal{P}$. Note that the edge modes respect $\mathcal{CP}$ symmetry even
though it is broken in the bulk, therefore the edge modes have only real
eigenfrequencies.

\emph{Conclusion and Discussion.} In conclusion, we propose a new class of
non-Hermitian photonics based on $\mathcal{CP}$ symmetry, which shares
certain similarity, but is dramatically different from the well-known $%
\mathcal{PT}$ -symmetric photonics. The physical realization of such $%
\mathcal{CP}$-symmetric photonics in chiral dielectrics opens a novel
pathway for engineering exotic hyperbolic materials with significant
applications. Our work may shed light on future experimental and theoretical
development of new non-Hermitian photonics.

Many important questions remain to be answered for the $\mathcal{CP}$%
-symmetric non-Hermitian photonics and here we list a few of them: \textit{i}%
) Can $\mathcal{CP}$ symmetry and its breaking be induced by means other
than chiral effects? \textit{ii}) Can $\mathcal{CP}$ symmetric photonics be
applied to periodic optical systems like photonic crystals or coupled
optical cavities? \textit{iii}) Are there other paths to non-Hermitian
photonics besides $\mathcal{PT}$ and $\mathcal{CP}$ symmetries? \textit{iv})
Finally, does such $\mathcal{CP}$-symmetric non-Hermitian physics exist in
physical systems other than photonics?

Acknowledgements: This work was supported by Air Force Office of Scientific
Research (FA9550-16-1-0387), National Science Foundation
(PHY-1505496,PHY-1806227), and Army Research Office (W911NF-17-1-0128). Z.
Li and Q. Gu acknowledge funding from UT Dallas faculty start-up funding.

%

\newpage \clearpage
\onecolumngrid
\appendix

\subsection{$\mathcal{C}$, $\mathcal{P}$, and $\mathcal{T}$ symmetries of
Maxwell equations}

\label{SecA} The condition for persevering $\mathcal{T}$ symmetry can be
simply obtained from matching $\mathcal{T}H_{M}\mathcal{T}^{-1}$ with $H_{M}$
since we already have $\mathcal{T}H_{P}\mathcal{T}^{-1}=H_{P}$. A simple
calculation shows $\mathcal{T}H_{M}\mathcal{T}^{-1}=\left(
\begin{array}{cc}
\epsilon ^{\ast } & i\gamma \\
-i\gamma & \mu ^{\ast }%
\end{array}%
\right) ,$ which yields the condition shown in the main text.

One can easily verify that $\mathcal{P}=-\sigma _{z}\otimes I_{3}$ satisfies
the following equations $\mathcal{P}\left(
\begin{array}{cc}
0 & \nabla \times \\
-\nabla \times & 0%
\end{array}%
\right) \mathcal{P}^{-1}=\left(
\begin{array}{cc}
0 & -\nabla \times \\
\nabla \times & 0%
\end{array}%
\right) $, $\mathcal{P}\left(
\begin{array}{c}
\bm{E} \\
\bm{H}%
\end{array}%
\right) =\left(
\begin{array}{c}
-\bm{E} \\
\bm{H}%
\end{array}%
\right) $, $\mathcal{P}\left(
\begin{array}{cc}
\epsilon & i\gamma \\
-i\gamma & \mu%
\end{array}%
\right) \mathcal{P}^{-1}=\left(
\begin{array}{cc}
\epsilon & -i\gamma \\
i\gamma & \mu%
\end{array}%
\right) $ and $|\mathcal{P}|=-1$. Therefore it fulfills all physical and
mathematical requirements of a parity symmetry operator. Since $H_{P}$ is
even under $\mathcal{P}$ and $H_{M}$ is even when $\gamma =0$, the system has even parity when $\gamma =0$, regardless of the forms of $\mu $ and $\epsilon $.

Photons are gauge bosons and their own antiparticles. Combining with the
anti-linear requirement, we can reasonably guess the charge conjungation $%
\mathcal{C}=e^{i\theta _{c}}I_{2}\otimes I_{3}K$. Furthermore, $\mathcal{C}%
^{2}=I_{6}$ because the system is expected to be invariant if the charge
conjugation operation is applied twice. These observations lead to $\mathcal{%
C}=\pm K$, where we have dropped the identity matrix.

These choices are self-consistent in the sense that $\mathcal{CPT}=1$ such
that Maxwell equations remain invariant under $\mathcal{CPT}$. The matrix
representations are valid only for Maxwell equations in the form of Equ.~\ref%
{Hamiltonian} and under the assumption of a spatially homogenous $H_{M}$.

\subsection{An explicit example of $\mathcal{CP}$ breaking}

\label{SecB} In the main text, we argue that the spontaneous breaking of $%
\mathcal{CP}$ symmetry may lead to complex eigenmodes, which can be
understood through the transformation of eigenmodes at $\pm \bm{k}$. If $(%
\mathcal{CP})\psi _{j,\bm{k}}=e^{i\theta_{j}}\psi _{j,-\bm{k}}$, there is a constraint on eigenfrequencies $\omega_{j,\bm{k}}=\omega _{j,-\bm{k}}^{\ast }$, which dictates real spectrum. Here, we show an example in both $\mathcal{%
CP}$-symmetric and $\mathcal{CP}$-broken regions by writing down the
eigenmodes explicitly. We take $\epsilon _{D}=4$, $\mu _{D}=1$, $\gamma =%
\text{diag}(0,0,\gamma_z)$ and $\bm{k}=\bm{1}$. The exceptional point is
then $\gamma _{z}^{c}=2$.

There are two non-zero and positive (in $\mathcal{CP}$-symmetric regime) solutions $\omega_{0,\pm\bm{k}}=(6-\gamma_z)/(2\gamma_{z,-})$ and $\omega_{1,\pm\bm{k}}=(6+\gamma_z)/(2\gamma_{z,+})$ with eigenstates (but we can only take either the positive or negative branches for a given system)
\begin{eqnarray}
\psi _{0,\pm\bm{k}} &=&((i(2-\gamma_z)\pm\gamma_{z,-})/8,
(i(2-\gamma_z)\mp\gamma_{z,-})/8,-i/2,(\gamma_z-2\pm i\gamma_{z,-})/4\
,(\gamma_z-2\mp i\gamma_{z,-})/4,1)^{T}, \\
\psi _{1,\pm\bm{k}} &=&(-(i(\gamma_z+2)\pm\gamma_{z,+})/8,
-(i(\gamma_z+2)\mp\gamma_{z,+})/8,i/2,-(\gamma_z+2\mp i\gamma_{z,+})/4,
-(\gamma_z+2\pm i\gamma_{z,+})/4,1)^{T},
\end{eqnarray}%
where $\gamma_{z,\pm}=\sqrt{(\gamma_z\pm 6)(\gamma_z\pm2)}$.

When $0<\gamma _{z}<\gamma _{z}^{c}$, $\gamma _{z,\pm }>0$. It is obvious
that $(\mathcal{CP})\psi _{j,\bm{k}}=e^{i\pi}\psi_{j,-\bm{k}}$
such that $\omega _{j,\bm{k}}=\omega _{j,\bm{k}}^{\ast}$, which is in the $\mathcal{CP}$-symmetric phase. When $\gamma _{z}>\gamma _{z}^{c}$, $\gamma
_{z,+}>0$ but $\gamma _{z,-}$ becomes purely imaginary. The condition $(%
\mathcal{CP})\psi _{1,\bm{k}}=e^{i\pi }\psi _{1,-\bm{k}}$
survives, so we still see a real eigenmode. The transformation for
the other one becomes $(\mathcal{CP})\psi _{1,\pm\bm{k}}=e^{i\pi}\psi _{1,\pm\bm{k}}$, which requires $\omega _{1,\pm\bm{k}}=-\omega _{1,\pm\bm{k}}^{\ast }$. The $\mathcal{CP}$ symmetry is partially broken by strong chiral effects and the
eigenspectra become complex.

\subsection{Hyperbolic bands from $\mathcal{CP}$ breaking by chiral effects}

\label{SecC}
\begin{figure}[t]
\includegraphics[width=0.5\textwidth]{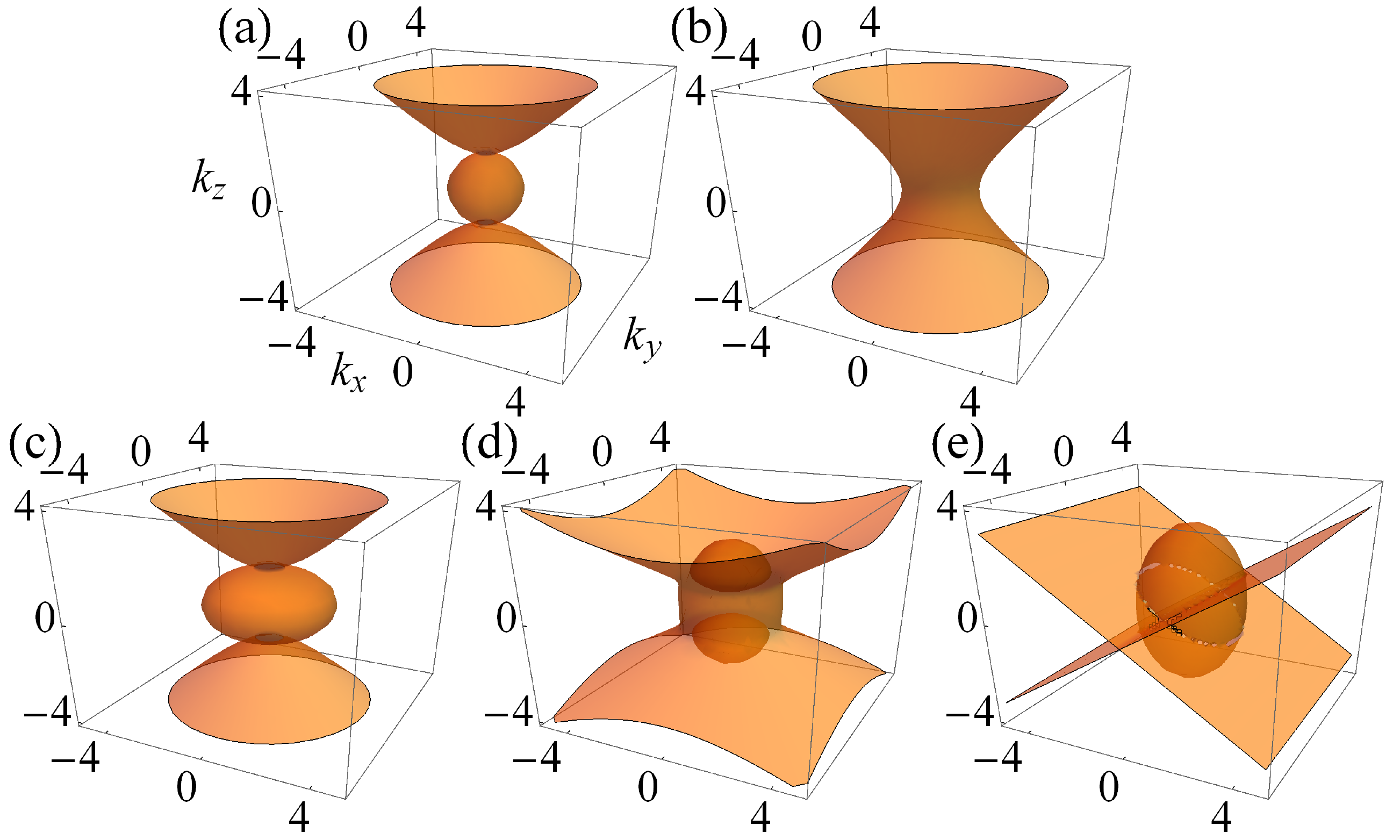} \centering
\caption{Typical EFSs at $\protect\omega =1$ for type-I (a) and type-II (b)
HMMs. We choose $\protect\epsilon =\text{diag}(2,2,-2)$ and $\protect%
\epsilon =\text{diag}(-2,-2,2)$ respectively. Corresponding realizations of
dispersion relations in strong chiral dielectrics are shown in (c) $\protect%
\epsilon =2I_{3}$, $\protect\gamma =\text{diag}(0,0,3)$, (d) $\protect%
\epsilon =\text{diag}(1,1,2)$, $\protect\gamma =\text{diag}(1.5,1.5,0)$ and
(e) $\protect\epsilon =\text{diag}(2,2,2)$ and $\protect\gamma =\text{diag}%
(2,\protect\sqrt{2},0)$. For all panels $\protect\mu =I_{3}$.}
\label{figS1}
\end{figure}

The EFS for pure dielectric materials are spheres (ellipsoids) in the
momentum space. However, the EFSs for HMMs are completely different as shown in
Fig.~\ref{figS1}(a,b), corresponding to type-I and type-II HMMs respectively
\cite{SmithDR2003,PoddubnyA2013}. A significant feature of these EFSs is
that they stretch to infinity in momentum space so that the material can support the
propagation of large $|\bm{k}|$ waves. Such hyperbolicity was thought to be
unique to HMMs \cite{PoddubnyA2013}. Quite surprisingly, similar dispersions
can also be realized through $\mathcal{CP}$-breaking induced by chiral
terms. The corresponding EFSs are plotted in Fig.~\ref{figS1}(c,d).

As observed in Fig.~\ref{figS1}(c), chiral term along one spatial direction
exceeding the exceptional point can render a dispersion relation mimicking
that of a type-I HMM. If the chiral terms along two spatial directions
exceed the exceptional points, the dispersion relation is similar as that
for a type-II HMM, as shown in Fig.~\ref{figS1}(d). Nevertheless, the
geometries are not exactly the same in the small $|\bm{k}|$ region because
we have two bounded EFSs surrounded by the hyperbolic one. While we have
presented similar results and some exotic hyperbolic dispersions in the main
text, Fig.~\ref{figS1}(e) offers an example where the system is $\mathcal{CP}
$-broken in one direction, and at the exceptional point in another direction.

In the following, we study the wave propagation in chiral media and show how
the hyperbolic dispersions emerge. We first consider a chiral isotropic
dielectric medium with constant scalar permittivity, permeability and chiral
effect $\gamma =\gamma _{D}I_{3}$ as defined in the main text. This simple
model is enough to gain insights into the systems and has been adopted
frequently in previous literatures \cite%
{LeknerJ1996,PendryJB2004,MonzonC2005,ChernR2013}. The resulting dispersion
relation
\begin{equation}
\omega =|\bm{k}|/|\sqrt{\epsilon _{D}\mu _{D}}\pm \gamma _{D}|.
\end{equation}%
For $\gamma _{D}=0$, the EFS is a two-fold degenerate sphere as expected and
the degeneracy comes from two different polarizations. For a finite chiral
strength, the degeneracy is lifted and there are always two spheres with
different radii in the momentum space except at $\gamma _{D}=\pm \sqrt{%
\epsilon _{D}\mu _{D}}$. It has been shown that for $|\gamma _{D}|>\sqrt{%
\epsilon _{D}\mu _{D}}$, there are negative refraction for proper incident
angles because the time-averaged Poynting vector $\langle \bm{S}\rangle _{t}$
is antiparallel to $\bm{k}$ \cite{ChernR2013}. This result is
straightforward by noticing that the refraction indexes are $n=\sqrt{%
\epsilon _{D}\mu _{D}}\pm \gamma _{D}$ for right- and left-handed circular
polarizations \cite{WangZ2016}. Nevertheless, the EFSs always remain
ellipsoids for both $\gamma _{D}<\sqrt{\epsilon _{D}\mu _{D}}$ and $\gamma
_{D}>\sqrt{\epsilon _{D}\mu _{D}}$ as a result of linear dispersion
relations. Due to the bounded EFSs, there exists a critical angle $\theta
_{c}$ in a scalar chiral medium, beyond which the incident waves with
certain polarizations are totally reflected. A scalar chiral term cannot
render spontaneous $\mathcal{CP}$ breaking, which requires strong
anisotropic chiral terms.

We now revisit the simple but instructive case in the main text with
slightly different $\epsilon =\text{diag}(\epsilon _{t}>0,\epsilon
_{t}>0,\epsilon _{z}>0)$, $\mu =1$ and $\gamma =\text{diag}(0,0,\gamma _{z})$%
. The dispersion relation becomes
\begin{equation}
(k_{t}^{2}+k_{z}^{2}-\epsilon _{t}\omega ^{2})(\epsilon
_{t}k_{t}^{2}+\epsilon _{z}k_{z}^{2}-\epsilon _{z}\epsilon _{t}\omega
^{2})=\gamma _{z}^{2}(k_{z}^{2}-\epsilon _{t}\omega ^{2})^{2},  \label{wave}
\end{equation}%
where $k_{t}^{2}=k_{x}^{2}+k_{y}^{2}$. Our previous analysis suggests that
there is an exceptional point $\gamma _{z}=\pm \sqrt{\epsilon _{z}}$ since $%
\gamma _{z}$ is decoupled from $k_{x}$ and $k_{y}$ in Equ.~\ref{wave}. For a
small $\gamma _{z}$, we have two distinguishable ellipsoids since the
degeneracies between different polarizations are lifted. As the chiral
strength increases, the inner EFS gets compressed along both $k_{x}$ and
$k_{y}$ directions and disappears at the exceptional point. After passing
the exceptional point, the EFSs are not two ellipsoids anymore because the
leading term $(\epsilon _{z}-\gamma _{z}^{2})\epsilon _{t}^{2}\omega ^{4}$
becomes negative, which mimics a type-I HMM. As a result, we observe type-I
hyperbolic dispersions in the $\mathcal{CP}$-broken phase since hyperbolic
bands must be non-Hermitian due to its metal character. Note that the two
degenerate points at $\bm{k}=(0,0,\pm k_{z})$ survives, but can be lifted by
anisotropy in permeability tensor $\epsilon _{x}\neq \epsilon _{y}$ or
additional chiral terms.

Although the chirality induced hyperbolic dispersion shares a similar EFS to that for HMM, it is impossible to obtain a homogenous model for a strong
chiral medium similar to that for a simple HMM. In a chiral medium, the
eigenmodes with different polarizations are not degenerate. The existence of chiral effects also break all spatial inversion symmetries while a pure HMM
preserves them. However, by comparing the wave equations in the
frequency-momentum domain, we can recast the chiral medium with only
non-zero $\gamma _{z}$ into a \textquotedblleft pure HMM\textquotedblright\
form
\begin{equation}
\epsilon _{eff}=\left( \epsilon _{t},\epsilon _{t},\epsilon _{z}-\gamma
_{z}^{2}(k_{z}^{2}-\epsilon _{t}\omega ^{2})/(k_{t}^{2}+k_{z}^{2}-\epsilon
_{t}\omega ^{2})\right) ,  \label{eff}
\end{equation}%
where the effective $\epsilon _{eff}$ along the $z$ direction depends on $%
k_{z}$ and $\omega $. Such projection cannot be applied to materials with
non-vanishing chiral effects along two directions because there are coupled
terms like $\gamma _{i}\gamma _{j},i\neq j$, which lead to the difference in
the geometries of EFSs (see Figs.~\ref{figS1}(b) and (d)).

\begin{figure}[t]
\includegraphics[width=1\textwidth]{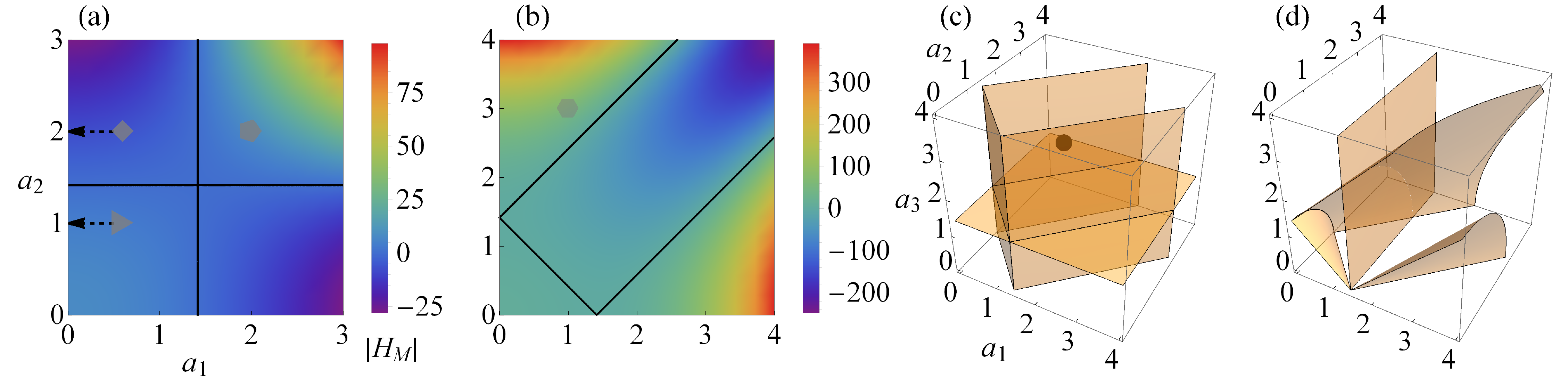} \centering
\caption{Phase diagram plotted by $\det |H_{M}|$ for different chiral terms.
(a) Density plot of $\det |H_{M}|$. The solid black curves denote zero
solutions (exceptional points). $\protect\gamma =\text{diag}(0,a1,a2)$. The
triangle, square and pentagon labels represent the parameters for panel
(a1), (a2) and (a3) in Fig.~\protect\ref{fig2} respectively. (b) Similar as
(a) but the chiral term is chosen as $\protect\gamma =\left(
\protect\begin{array}{ccc}
a1 & a2 & 0\protect \\
a2 & a1 & 0\protect \\
0 & 0 & 0%
\protect\end{array}%
\right) $. The hexagon corresponds to Fig.~\protect\ref{fig2}(a4). (c,d)
Phase boundary for chiral terms $\protect\gamma =\left(
\protect\begin{array}{ccc}
a1 & a2 & 0\protect \\
a2 & a1 & 0\protect \\
0 & 0 & a3%
\protect\end{array}%
\right) $ and $\protect\gamma =\left(
\protect\begin{array}{ccc}
a1 & a2 & a3\protect \\
a2 & a1 & 0\protect \\
a3 & 0 & a1%
\protect\end{array}%
\right) $. The black ball gives the parameters used in Fig.~\protect\ref%
{fig2}(a5).}
\label{figS2}
\end{figure}

Although exceptional points along three spatial directions are decoupled
when $\gamma $ and $\epsilon $ (or $\mu $) commute, the situation can be
much more complicated if there are non-diagonal terms in the chiral tensor.
When there are only diagonal chiral terms, the exceptional points form two
intersecting exceptional lines that separate the phase diagram into four
parts (Fig.~\ref{figS2}(a)). The phase boundary changes when a non-diagonal
term is considered (Fig.~\ref{figS2}(b)), which is accompanied with some
exotic hyperbolic dispersions. In 3D parameter spaces, there exist
exceptional surfaces (Fig.~\ref{figS2}(c,d)). We see the complex chiral
configurations lead to hyperbolic dispersions that cannot be realized in
regular HMMs.

\subsection{Application: all-angle polarization-dependent beam splitter}

\label{SecD} As we discussed before, a scalar chiral term could render
negative refraction but only for a small range of incident angle. This effect is
illustrated in Fig.~\ref{fig3}(a) as the solid curves, where we choose $x$-$%
y $ plane as the plane of incidence, $y$ axis as norm and the incident beam
comes from the right side of the norm. The chiral medium locates at $y<0$
and the region $y\geq 0$ is vacuum. The incident and refraction angles are $%
\theta _{i}$ and $\theta _{t}$ respectively. The negatively refracted beam
disappears when the incident angle exceeds a critical value $\theta _{c}\sim
36^{\circ }$, which can be attributed to bounded EFSs. As a comparison, we
also plot the $\theta _{i}$-$\theta _{t}$ relation for a pure dielectric as
the dashed curve. There is only one single curve due to the lack of
birefringence effect.

Thanks to the hyperbolic dispersion, all-angle negative refraction can be
realized in HMMs \cite{LiuY2008}, which is polarization independent. In the $%
\mathcal{CP}$-broken region, the hyperbolic dispersion allows the
realization of all-angle polarization-sensitive birefringence and negative
refraction. We consider a dielectric with chiral vector $\gamma =\text{diag}%
(0,2,0)$, which is in the $\mathcal{CP}$-broken phase with a type-I
hyperbolic dispersion. In Fig.~\ref{fig3}(a1), we see that the negative
refraction indeed happens for arbitrary incident angles. We further confirm
these results through COMSOL simulations, which are plotted in Fig.~\ref%
{fig3}(a2,a3). There is no negative refraction when $\theta _{i}>\theta _{c}$ in
the $\mathcal{CP}$-symmetric phase. This effect can be used to engineer an
all-angle polarization beam splitter.

\subsection{Application: enhanced spontaneous emissions for laser engineering%
}

\label{SecE} Spontaneous emissions in chiral media have been studied in
various situations and exhibit many interesting phenomena. Previous studies
have mainly focused on isotropic $\gamma $. The $\mathcal{CP}$ breaking
through strong anisotropic $\gamma $ may significantly enhance the
spontaneous emissions of a dielectric continuum with both broader bandwidth
and stronger Purcell effect due to its hyperbolic bands \cite{FerrariL2014}.

Through Fermi's golden rule, we can easily see that the radiative decay rate
is generally proportional to the photonic density of states $\rho (\omega
)=\sum_{\sigma ,\bm{k}}\delta (\omega _{\sigma ,\bm{k}}-\omega )$, where the
summation goes over all polarizations $\sigma $ and momenta $\bm{k}$. The
density of state is proportional to the area of EFSs at $\omega _{\sigma ,%
\bm{k}}=\omega $, which is a small finite value for dielectrics but diverges
for HMMs. Note that it does not diverge in real physical systems due to
finite-size effects and corrections of effective medium theory for large $%
\bm{k}$ states \cite{PoddubnyA2013}. Based on the above arguments, a
hyperbolic dispersion would render a larger radiative decay rate and thus, a stronger Purcell effect. The Purcell effect characterizes the enhancement or inhibition of spontaneous emission in a system with respect to free space. For most nanophotonic applications, a stronger Purcell effect (i.e. a larger Purcell factor) is desired.

With the physical understanding, we expect to observe a jump of Purcell
factor in chiral medium at a given wavelength $k$ upon crossing the
exceptional point and entering the $\mathcal{CP}$-broken regions. Moreover,
the Purcell effect should be much stronger in these cases with exotic
hyperbolic dispersions (see Fig.~\ref{fig2}(a4,a5)) as they are expected to
have larger photonic density of state due to the lack of bounded EFSs.

These expectations are further confirmed by COMSOL simulations as shown in
Fig.~\ref{fig3}(b). We consider a chiral medium with scale 100nm$\times $%
100nm$\times $200nm and dielectric constants $\epsilon =2$ and $\mu =1$. The
chiral terms are $\gamma =(0,0,\gamma _{0})$ and $\gamma =(0,\gamma
_{0},\gamma _{0})$ for each panel. We set $\gamma _{0}=0,1.4,1.5,2$ and $3$
for the curves in blue, khaki, green, red and purple colors, respectively.
An electric dipole source in vacuum is placed 10nm above the chiral medium
in $x$-$y$ plane and the plotted Purcell factor is averaged through three
dipolar configurations along three spatial directions \cite{LuD2014}. In the
$\mathcal{CP}$-symmetric regime, the emission strength is essentially
unchanged. A substantial enhancement is observed after $\gamma _{0}$ passes
the exceptional point and the system enters the $\mathcal{CP}$-broken regime.

\subsection{Application: topological photonics}

\label{SecF} While the current research on topological photonics have
focused on periodic systems like photonic crystals, it was recently shown
that a continuous HMM is also topologically non-trivial under proper
symmetry-breaking fields \cite{GaoW2015,ChernRL2017,HouJ2018}. In this work,
we characterize the band structure of chiral medium through the methods
developed in \cite{HouJ2018}.

The band structure of a $\mathcal{CP}$-symmetric chiral medium is plotted in
Fig.~\ref{fig2}(b) and there cannot be any finite gap no matter how the
bands are projected. In this sense, although we may have a strictly
quantized Chern number for each bands, this is a trivial phase without any
surface state. With increasing $\gamma $ along $z$ direction, the lower band
approaches to the zero plane. Upon passing the $\mathcal{CP}$ breaking
point, the upper and lower bands only have one degenerate point in the
subspace $k_{z}\neq 0$ as shown in Fig.~\ref{fig2}(b). Such a degeneracy can
be easily lifted by anisotropy of $\epsilon $ or gyromagnetic effects. After
the band gap is opened, the two bands are now topologically non-trivial with
a quantized band Chern number and may support chiral surface wave. The only
degeneracy in the momentum space is the origin, which is a triply-degenerate
point with a topological charge 2 \cite{HouJ2018}.

In Fig.~\ref{fig3}(c1), we show the projected 2D band together with the
chiral surface wave solutions, where we choose $\epsilon =\text{diag}(3,2,1)$%
, $\mu =I_{3}$ and $\gamma =\text{diag}(0,0,0.5)$. The existence of the
chiral surface wave is also confirmed by the COMSOL simulations shown in
Fig.~\ref{fig3}(c2). The surface wave has different chiralities at $%
k_{z}=\pm 1$ due to the non-vanishing charge of triply-degenerate point
(breaking $\mathcal{P}$ symmetry). Our results indicate that the
non-Hermitian band theory formulated from Maxwell equations for HMMs \cite%
{HouJ2018} can also be generalized to chiral dielectric materials.

\end{document}